# Long-distance frequency transfer over an urban fiber link using optical phase stabilization

H. Jiang,[1] F. Kéfélian,[2] S. Crane,[1,3] O. Lopez,[2] M. Lours,[1] J. Millo,[1] D. Holleville,[1]

P. Lemonde,[1] Ch. Chardonnet,[2] A. Amy-Klein,[2] G. Santarelli[1,*]

[1]LNE-SYRTE, Observatoire de Paris, CNRS, 61 avenue de l'Observatoire, 75014 Paris, France

[2]Laboratoire de Physique des Lasers (LPL), UMR 7538, CNRS and Université Paris 13, 99 av. J.-B. Clément, 93430 Villetaneuse, France

[3]Permanent address: United States Naval Observatory, 3450 Massachusetts Avenue NW, Washington, DC 20392, USA

[*]Corresponding author: giorgio.santarelli@obspm.fr

We transferred the frequency of an ultra-stable laser over 86 km of urban fiber. The link is composed of two cascaded 43-km fibers connecting two laboratories, LNE-SYRTE and LPL in Paris area. In an effort to realistically demonstrate a link of 172 km without using spooled fiber extensions, we implemented a recirculation loop to double the length of the urban fiber link. The link is fed with a 1542-nm cavity stabilized fiber laser having a sub-Hz linewidth. The fiber-induced phase noise is measured and cancelled with an all fiber-based interferometer using commercial off the shelf pigtailed telecommunication components. The compensated link shows an Allan deviation of a few $10^{-16}$ at one second and a few $10^{-19}$ at 10,000 seconds.

# 1. INTRODUCTION

The transfer of ultra-stable frequencies between distant laboratories is required by many applications in time and frequency metrology, fundamental physics, particle accelerators and astrophysics. Remote clock comparisons are currently performed using satellites, either directly by Two-Way Satellite Time and Frequency Transfer, or indirectly through the Global Positioning System carrier phase measurement. However, both methods are limited by instability of $10^{-15}$ after one day of averaging time [1] and are consequently insufficient to transfer modern cold atom microwave frequency standards having demonstrated frequency stability of a few $10^{-16}$ at one day [2].

Progress on satellite-based links may lower the noise floor to the $10^{-16}$ range at one day and advanced space missions such as "ACES" (Atomic Clocks Ensemble in Space) [3] and "T2L2" [4] are being designed to make comparisons in the high $10^{-17}$ at one day. However, even if achieved, such performance would be insufficient for the next generation of optical clocks. Optical clocks using a trapped single ion or cold atoms confined in lattice are expected to reach instability level of $10^{-17}$ or better at one day and will consequently require even more stable frequency transfer systems [5,6]. Moreover the assets of the accuracy and stability of the optical clock are critically related to an efficient way to compare remote clocks with short averaging time. Beyond metrology, high-resolution clock comparison is essential for advanced tests in fundamental physics, such as tests of the fundamental constants stability [7].

To overcome current space link limitations, the transmission of frequency standards over optical fiber has been investigated for several years [8]. This technique takes advantage of the fact that fiber has low attenuation, high reliability and the potential for phase noise cancellation.

Microwave frequency transmission using amplitude modulation of an optical carrier demonstrated instability as low as $2\times10^{-18}$ at one day over 86 km [9,10,11]. Direct optical frequency transfer [12,13,14,15] can provide even better stability and be extended to greater distance. Indeed, optical frequency transfer is less sensitive to link attenuation, due to heterodyne detection. In addition, the higher carrier frequency gives much better resolution for measuring link-induced phase noise. In 2007, two pioneering experiments of optical frequency transfer over a fiber link of more than 200 km were reported [13,14]. Both experiments used fiber spool extension of an urban link and demonstrated the feasibility of a full optical link with instability in the $10^{-18}$ range. Since last year several German research laboratories have been connected by 1000 km of dedicated fiber from the national research network and the stabilization of the link is under development [16]. These are the first milestones towards continental scale fiber links.

In this paper, we report the transmission of a sub-Hz linewidth optical frequency reference over an 86 km length, which we doubled in a second step to an extended length of 172 km through a recirculation loop. After the description of the cavity-stabilized laser used to test the link, we present the scheme of the link-induced phase noise compensation system. The frequency stability performance of an 86-km urban fiber link is then reported and, finally, the link length is extended to 172 km and characterized.

## 2. LASER SOURCE

Two identical ultra low noise laser sources have been developed using 1542-nm 1-2-kHz linewidth commercial fiber lasers stabilized on two identical ultra-stable cavities by the Pound-Drever-Hall method [17]. The cavity consists of a 10-cm ULE spacer and two optically contacted ULE mirrors, giving a measured finesse of ~ 800,000. The cavity, specially designed to minimize the vibration sensitivity in all spatial directions [18], is mounted horizontally and sits on four Viton pads. It is placed in a vacuum chamber with pressure below $10^{-7}$ mbar and the entire system is on a vibration-isolation platform in an acoustical isolation box. About 2 μW of optical power, including ~ 30% in the phase modulation sidebands, are typically sent onto the cavity with a coupling efficiency better than 60%. A 350-kHz bandwidth locking is achieved using a double-pass acousto-optic modulator with a cat's eye retroreflector. The stabilized output of the laser can provide up to 10 mW of optical power.

To measure the laser frequency noise power spectral density, the two independent cavity stabilized lasers are mixed on a photodetector, and the down-converted beat-note signal is measured with a dynamic signal analyzer after frequency-to-voltage conversion. The phase noise power spectral density ~~versus Fourier frequency~~ of the laser is presented on Fig. 1. Except for some spurious peaks between 40 Hz and 100 Hz, it is largely below the phase noise of a 1-Hz linewidth white frequency noise laser (dashed line on Fig. 1). The integrated phase noise from 1 Hz to 10 kHz is below 0.2 rad rms (root mean square). The fractional frequency instability (Allan deviation) was also calculated from the frequency counted beat-note signal and was found less than $2\times10^{-15}$ at 1 s and $10^{-14}$ at 100 s after a 0.3-Hz/s drift was removed.

## 3. COMPENSATED LINK SETUP

Fig. 2 shows the scheme of the compensated link based on the principle first described in [19]. The ultra-stable laser light is divided into two parts using a fiber coupler. One arm provides the reference signal for stability measurement and fiber-induced phase noise compensation, while the other arm is connected to the link through an optical circulator (OC) followed by acousto-optic modulator AOM1 (with frequency $f_1 \approx 40$ MHz). To compensate for the phase noise $\phi_p$ accumulated along the fiber due to acoustical, mechanical and thermal perturbations, part of the signal at the remote end is retraced back to the link through an optical circulator, after frequency shifting by acousto-optic modulator AOM2 (with frequency $f_2=70$ MHz). This return signal, which passes twice through the link and experiences a phase noise $2\phi_p$, is mixed at the local end with the reference signal on photodiode PD1. The beat note at frequency $2f_1+f_2$ is phase locked to a stable RF synthesizer using AOM1 driven by a voltage controlled oscillator. The phase-lock loop applies the correction $\phi_c=-\phi_p$ to the AOM1 frequency $f_1$, thus to the optical signal phase and consequently actively cancels the fiber-induced phase noise at the remote end of the fiber link. To magnify the dynamic range of the servo loop and hence improve its robustness, a digital frequency divider by 40 has been used just ahead of the phase detector. The optical frequency (phase) instability of the link is defined as the difference between the local and remote end optical frequencies (phases). It is measured on the beat-note at frequency $f_1+f_2$ provided by mixing the single trip and the reference optical signals on photodiode PD2. Two polarization controllers are employed for optimizing the beat-note signal amplitudes. The compensation system is entirely fibered and uses only commercial off the shelf pigtailed telecommunications components.

## 4. TRANSMISSION OVER 86 KM

LPL and LNE-SYRTE, located in Paris area, are linked by a pair of 43-km telecommunication fibers. Each fiber is composed of various sections of buried cables of the metropolitan network spliced together. The attenuation of each fiber is about 12 dB. By connecting the two ends of the fibers at LPL, the link length is extended to 86 km with local end and remote end both located in LNE-SYRTE.

Fig. 3 shows the 86-km link optical phase noise power spectral density ~~versus Fourier frequency~~, without and with compensation. The beat note provided by PD2 is frequency divided by 40 and down mixed to DC, the phase noise power spectral density is then measured with a dynamic signal analyzer. It exhibits a roughly $1/f^2$ roll-off from 1 Hz to 100 Hz with a wide bump between 10 Hz and 100 Hz, and $1/f^4$ behavior above 100 Hz. The integrated phase noise from 1 Hz to 1 kHz is equal to 19.2 rad and 0.4 rad (rms), without and with compensation respectively (corresponding to 15 fs and 0.3 fs rms timing jitter). Beyond 1 kHz, the phase noise rolls down to the measurement system floor and is negligible. The 0.4 rad (rms) corresponds to 85% of the optical power in the carrier. The phase noise of the transmitted laser is only slightly degraded by the transfer and the laser coherence time is therefore preserved at the output of the compensated link. This is an important point for applications requiring distribution of narrow linewidth sources.

Laser phase noise can corrupt the link-induced phase noise measurement. Indeed, on PD2 are mixed two optical signals coming from the same source but delayed by 430-μs (due to the 86-km propagation in the fiber). This generates on the beat-note signal a delayed self-heterodyne interferometric phase noise [20] due to the laser phase noise, in addition to the link-induced phase noise to be compensated. This additional phase noise contribution can be calculated from the laser phase noise (shown on Fig. 1) and is displayed in Fig. 3 (c). It shows that the laser phase noise is sufficiently low and does not limit the performance of the stabilization.

Fig. 4 displays the experimental fiber-induced optical phase noise rejection for the 86-km link versus Fourier frequency calculated by dividing the phase noise power spectral densities of the compensated and uncompensated links. We have performed the Laplace domain analysis of the compensation scheme and obtain an analytical expression of the rejection transfer function (black dashed line on Fig. 4) [21]. This analysis reveals that the optimum proportional gain is about the inverse of the link single trip time (2300 for an 86-km link) and that the rejection frequency bandwidth is limited to $1/4t_{trip}$ ($\sim 600$ Hz) where $t_{trip}$ is the one-way trip delay in agreement with [14]. Moreover, this analysis shows that the integrator used in the phase-lock loop improves rejection only in a limited frequency range. Indeed, as already pointed out in [14], at low frequency the rejection is limited by the link delay and scales as $(f.t_{trip})^2$ (red dotted line on Fig. 4) independently of the loop gain. Our calculations are in good agreement with the measurements as shown on Fig. 4.

For long term frequency stability characterization, the Allan deviation is the most common tool and is typically obtained using frequency counter measurements. The original Allan deviation defined in [22] can only be calculated from frequency samples obtained with a classical, or "π-type", counter using a uniform average over the measurement gate time. Indeed, modern enhanced-resolution, or "Λ-type", counters have been shown to lead [23], not to the classical Allan deviation, but to a quantity proportional to the modified Allan deviation [24]. Consequently, we measure the classical Allan deviation of the fractional frequency instability introduced by the link with a four-channel π-type frequency counter. This counter is dead-time free in order to avoid bias in the calculation of the Allan deviation [25]. We measure simultaneously the frequency instability of the compensated link and the frequency instability of the compensation signal (at the voltage controlled oscillator output), which represents the free running fiber frequency noise. The beat-note between the remote end and local end optical signals provided by PD2 is frequency divided by 40 and band limited by a tracking filter based on a low noise quartz oscillator giving a measurement bandwidth of ~10 Hz. This bandwidth is larger than the inverse of the gate time.

Fig. 5 shows the fractional frequency overlapping Allan deviation of the 86-km link with and without compensation for several days of continuous operation. The Allan deviation scales down with a $1/\tau$ slope from 1 s to 100 s. The bump at 250 s, which corresponds to the half cycle time of the air conditioning system, is due to thermal effects in the sections of the interferometric system that are not actively compensated. After 1000 s of averaging time, the Allan deviation levels off in the low $10^{-19}$ range probably also limited by the uncompensated section of the system. Without filtering, the Allan deviation is 10 times higher. This is consistent with the fact that the Allan deviation of a white phase noise is proportional to the square root of the noise bandwidth.

## 5. TRANSMISSION OVER 172 KM

Previous extended link results have been demonstrated by adding spooled fiber to an urban link [13,14,15]. To simulate a longer link with a more realistic phase noise, we devised a new scheme, represented in Fig. 6, to pass twice through the 86-km link leading to a full urban link of 172 km. The optical signal is fed into the 86-km link through an optical coupler. At the end of the 86-km link, a "frequency shifter mirror" is connected. This "mirror" consists of a unidirectional loop based on an optical circulator and an acousto-optic modulator. An Erbium doped fiber amplifier (EDFA1) with 17-dB gain is also implemented in order to compensate for losses due to recirculation (12 dB due to the double pass combiner losses, 5 dB due to the optical circulator and AOM2 losses). As a result, this is equivalent to a 172-km span link without intermediate optical amplifier and the resulting phase noise is similar to the noise of a real 172-km telecom network link. A second optical amplifier (EDFA2) was added at the remote end to amplify the return signal. Temperature variations in EDFAs induce phase fluctuations, which degrade the long-term frequency stability. In order to correctly compensate for this effect, the output optical signal should pass once through the EDFAs when the round trip signal should pass twice. This is naturally the case for EDFA1, but not for EDFA2. This is overcome by adding a recirculation loop around EDFA2. To easily identify the output and round trip signals AOM3 is inserted in this loop. At the output of PD1 the beat-note signal at frequency $2(f_1+f_2)+f_3+f_4$ is appropriate for link compensation while at the output of PD2 the beat-note signal at $f_1+f_2$ is used to derive the link stability ($f_i$ is the frequency of the AOM $i$). This set-up could be simplified by using a bi-directional EDFA.

Fig. 7 displays the 172-km link phase noise power spectral density, without and with compensation. As expected, the correction bandwidth is half the one obtained for the 86-km link due to a double roundtrip delay. Moreover, at low frequency, the fiber-induced phase noise rejection is 4 times lower than with the 86-km link, in agreement with the delay-limitation effect discussed in section 4. The integrated phase noise from 1 Hz to 1 kHz is equal to 53 rad and 2.4 rad (rms), without and with compensation respectively.

Fig. 8 shows the fractional frequency instability of the 172-km link. The Allan deviation is about $4\times10^{-16}$ at 1 second and in the range of $10^{-19}$ at 1 hour with a 10-Hz measurement bandwidth. The system floor is measured by replacing the urban fiber with an optical attenuator having the equivalent attenuation.

In addition to data obtained with π-type counter we have used a Λ-type counter, without the 10-Hz filter, to allow comparison with the results of [14,15] obtained with 76 km of urban fiber, 175 km of spooled fiber and four in-line EDFA. To avoid dead-time, the gate time of the counter was set equal to the averaging time for averaging time ≤ 100 s. Data obtained with Λ-type counter are shown with green star points on Fig. 8. They are better than with π-type counter, due to an additional filtering of the phase noise. The Allan deviation is calculated for every averaging time with the classical formula and not with the modified Allan deviation (as in reference [14,15] for averaging time >10s). Indeed, when original Allan deviation formula is used, Λ-type counters report directly a quantity similar to the modified Allan deviation. For comparison, the use of the modified Allan deviation would lead in our case to a reduction factor of 2/3 on the Allan deviation for averaging time above 100 s. Results are found similar to the data presented in [14,15]. Data obtained with longer averaging times show that frequency instability keeps falling off until 5000 s and reach $3\times10^{-19}$ with the Λ-type counter at 3000 s.

Comparison between the two configurations is not straightforward. First, the link noise process is not time stationary due to urban environmental fluctuations. Moreover when the length of the fiber is virtually doubled by the recirculation technique the phase noise power spectral density of the free running link is twice the one of a single pass double length fiber, as detailed below.

Under the assumption that the fiber phase noise is uniformly spatially distributed, the phase noise power spectral density is proportional to the length of the fiber. However, with the recirculation technique the laser wave experiences twice the phase fluctuation at each point of the fiber. These two contributions of each point are correlated for times longer than the single span fiber delay $t_{fiber}$, they therefore add up coherently. Consequently, for Fourier frequencies below $1/4t_{fiber}$, the phase noise power spectral density of a $2L$-km link realized by recirculation is expected to be four times that of an $L$-km fiber. One can therefore anticipate that the phase noise power spectral density for the recirculated link is twice larger than would be expected for a real link of 172-km. Consequently the stability results obtained for the recirculated link should be considered as an upper bound for a real link of the same length.

# 6. DISCUSSION

It is conceivable to extend the length of the link up to continental scales, 1000 km or more. In a single segment approach however, major difficulties will arise. Firstly, fiber attenuation will make mandatory in-line bidirectional amplifiers in order to provide sufficient power at the output and allow phase locking at the local end. As discussed in section 5, bidirectional amplifiers phase noise is cancelled within the control bandwidth. However, when cascading amplifiers, amplified spontaneous emission can be very large and will require the use of narrow band optical filters to overcome a significant degradation of the signal-to-noise ratio at both ends [14, 15]. Secondly, the reduced bandwidth of the servo-loop and the reduced noise rejection level both due to the delay effect (See Section 4) will degrade the performance of the transmission. We have empirically observed that the phase noise exhibits a $1/f^2$ dependence in the free running link. Assuming a linear dependence of the phase noise power on the length of the link $L$ Allan deviation scales as $L^{3/2}$ [14,15]. One could hence expect an Allan deviation below $10^{-14}$ at one second for a 1000-km single segment link in a 10-Hz measurement bandwidth, averaging down as $1/\tau$ to the noise floor of the system.

An alternative approach is to split the long distance link in shorter compensated segments using intermediate stations. Each one will achieve three functions, to send back part of the received signal to the previous station, to amplify (and filter if necessary) the received signal, to compensate the phase noise induced by the following segment. The maximum distance of each segment will depend on the fiber phase noise distribution, attenuation and on the total link length. This multiple segment approach allows for an increased correction bandwidth and enhances the resolution of the system. This would enable much better performance down to a level where the coherence of the laser itself can be transferred, as demonstrated in section 4.

Beside the limitations due to the length of the link an other possible source of transmission degradation is related to spurious back reflections which always occurs in a fiber system. They are due to connector interfaces (typically -40dB to -60 dB), Rayleigh backscattering (typically -40 dB with SMF fiber) and splicing points. These effects can be enhanced when online bidirectional amplifiers are used. However, optical frequency transfer over fiber is not sensitive to single back-reflection as back-reflected waves do not receive the correct frequency offsets from the AOMs in our scheme and are consequently easily discriminated. This is not the case when double back-reflections are considered. Double back-reflected waves lead to additional phase and amplitude noise that is proportional to the fraction of the optical field double back-reflected. Since double back-reflections occur identically on the way there and the way back, the double back-reflection optical phase noise are also cancelled by the compensation system. Consequently, back-reflections, even strong, do not limit the stability of the link.

# 7. CONCLUSION

We have demonstrated a long-distance link for metrological optical frequency transfer using optical phase stabilization over a dedicated fiber that lies buried beneath the urban environment of Paris, France. Frequency transfer was demonstrated with instability of $1.5 \times 10^{-16}$ at 1 s and integrates down to the range of $10^{-19}$ at 500 s over 86 km. We have simulated a link at twice this distance via recirculation in the fiber to represent the phase noise of a 172-km urban link. It shows instability of $4 \times 10^{-16}$ at 1 s and a few $10^{-19}$ at 10,000 s, which is better than the anticipated performance of optical clocks.

To broaden the use of this technique to more users and reach longer distances, an appealing solution is to use the existing research, educational as well as commercial optical communication networks that connect a large number of physics laboratories. This will require addressing several difficulties. For instance, such a link will have to coexist with telecommunication traffic via the use of wavelength division multiplexing. Another crucial point is that fibers in classical optical telecommunication networks are used in a unidirectional way, whereas the compensation technique will require bi-directional operation. Yet one more significant hurdle being that restricted access to the fiber network infrastructure will require the design of an ultra-stable link that functions in a virtually flawless fashion. We are currently developing solutions to make compatible the frequency transfer technique and the optical communication networks, in particular to bypass unidirectional optical amplifiers and to allow the transparent coexistence of the ultrastable signal and the data modulated signal. This is the next challenging step towards a network for ultra stable frequency transfer.

# ACKNOWLEDGMENTS

We acknowledge funding support from the Agence Nationale de la Recherche (ANR BLAN06-3_144016). SYRTE is a Unité Mixte de Recherche of CNRS, Observatoire de Paris and UPMC.

## REFERENCES

1.    A. Bauch, J. Achkar, S. Bize, D. Calonico, R. Dach, R. Hlavac, L.Lorini, T. Parker, G. Petit, D. Piester, K. Szymaniec and P. Uhrich,"Comparison between frequency standards in Europe and the USA at the $10^{-15}$ uncertainty level", Metrologia, **43**, 109-120 (2006).

2.    C. Vian, P. Rosenbusch, H. Marion, S. Bize, L. Cacciapuoti, S. Zhang, M. Abgrall, D. Chambon, I. Maksimovic, P. Laurent, G. Santarelli, A. Clairon, A. Luiten, M. Tobar and C. Salomon, "BNM-SYRTE fountains: recent results" IEEE Transactions on Instrumentation and Measurement, **54**, 833- 836 (2005)

3.    L. Cacciapuoti, N. Dimarcq and C. Salomon, "Atomic Clock Ensemble in Space: Scientific Objectives and Mission Status", Nuclear Physics B **166**, 303-306 (2007)

4.    E. Samain P. Vrancken, J. Weick, P. Guillemot, "T2L2 Flight Model Metrological Performances" in *Proceedings of IEEE International Frequency Control Symposium and European Frequency and Time Forum.* (IEEE, 2007) pp. 1291-1294

5.    A. D. Ludlow, T. Zelevinsky, G. K. Campbell, S. Blatt, M. M. Boyd, M. H. G. de Miranda, M. J. Martin, J. W. Thomsen, S. M. Foreman, Jun Ye, T. M. Fortier, J. E. Stalnaker, S. A. Diddams, Y. Le Coq, Z. W. Barber, N. Poli, N. D. Lemke, K. M. Beck, and C. W. Oates, "Sr lattice clock at $1 \times 10^{-16}$ fractional uncertainty by remote optical evaluation with a Ca clock," Science **319**, 1805-1808 (2008).

6.    T. Rosenband, D. B. Hume, P. O. Schmidt, C. W. Chou, A. Brusch, L. Lorini, W. H. Oskay, R. E. Drullinger, T. M. Fortier, J. E. Stalnaker, S. A. Diddams, W. C. Swann, N. R. Newbury, W. M. Itano, D. J. Wineland, and J. C. Bergquist, "Frequency ratio of $Al^+$ and $Hg^+$ single-ion optical clocks; metrology at the 17th decimal place," Science **319**, 1808-1812 (2008).

7.    See for instance S.G. Karshenboim, Can. J. Phys. **83**, 767-811, (2005).

8.      S. M. Foreman, K. W. Holman, D. D. Hudson, D. J. Jones, and J. Ye, "Remote transfer of ultrastable frequency references via fiber networks", Rev. Sci. Instrum. **78**, 021101 (2007)

9.      C. Daussy, O. Lopez, A. Amy-Klein, A. Goncharov , M. Guinet, C. Chardonnet, F. Narbonneau, M. Lours, D. Chambon, S. Bize, A. Clairon, G. Santarelli, M.E. Tobar and A.N. Luiten, "Long-Distance Frequency Dissemination with a Resolution of $10^{-17}$", Phys. Rev. Lett. **94**, 203904 (2005).

10.     F. Narbonneau, M. Lours, S. Bize, A. Clairon, G. Santarelli, O. Lopez, C. Daussy, A. Amy-Klein, C. Chardonnet, "High resolution frequency standard dissemination via optical fiber metropolitan network", Rev. Sci. Instrum. **77**, 064701 (2006).

11.     O. Lopez, A. Amy-Klein, C. Daussy, Ch. Chardonnet, F. Narbonneau, M. Lours, and G. Santarelli, "86-km optical link with a resolution of $2\times10^{-18}$ for RF frequency transfer", Eur. Phys. J. D **48**, 35-41 (2008).

12.     S.M. Foreman, A.D. Ludlow, M.H.G. de Miranda, J. Stalnaker, S.A. Diddams, and J. Ye, "Coherent Optical Phase Transfer over a 32-km Fiber with 1 s Instability at $10^{-17}$", Phys. Rev.Lett. **99**, 153601 (2007).

13.     G. Grosche, B. Lipphardt, H. Schnatz, G. Santarelli, P. Lemonde, S. Bize, M. Lours, F. Narbonneau, A. Clairon, O. Lopez, A. Amy-Klein, Ch. Chardonnet , "Transmission of an Optical Carrier Frequency over a Telecommunication Fiber Link", in *Conference on Lasers and Electro-Optics/Quantum Electronics and Laser Science Conference and Photonic Applications Systems Technologies*, OSA Technical Digest (CD) (Optical Society of America, 2008), paper CMKK1.

14.     R. Newbury, P. A. Williams, W. C. Swann, "Coherent transfer of an optical carrier over 251 km", Opt. Lett. **32**, 3056-3058 (2007).

15.     P.A. Williams, W.C Swann, and N.R. Newbury, "High-stability transfer of an optical frequency over long fiber-optic links", J. Opt. Soc. Am. B **25**, 1284-1293 (2008).

16.     Harald Schnatz, PTB, private communication.

17.  R. W. P. Drever, J.L. Hall, F.V. Kowalski, J. Hough, G.M. Ford, A.J. Munley and H. Ward, "Laser phase and frequency stabilization using an optical resonator", Appl. Phys. B **31**, 97-105 (1983).

18.  J. Millo, S. Dawkins, R. Chicireanu, D. Varela, C. Mandache, D. Holleville, M. Lours, S. Bize, P. Lemonde, G. Santarelli, "Stable Optical Cavities Design and Experiments at the LNE-SYRTE", Proceedings of the 22nd European Frequency and Time Forum, Toulouse, France, 23-25 April 2008.

19.  L.-S. Ma, P. Jungner, J. Ye and J. L. Hall, "Delivering the same optical frequency at two places: accurate cancellation of phase noise introduced by an optical fiber or other time-varying path", Opt. Lett. **19**, 1777-1779 (1994).

20.  L. E. Richter, H. I. Mandelberg, M. S. Kruger, and P. A. McGrath, "Linewidth determination from self-heterodyne measurements with subcoherence delay times" IEEE Journal of Quantum Electronics, **22**, 2070-2074 (1986).

21.  H. Jiang, F. Kéfélian, S. Crane, O. Lopez, M. Lours, J. Millo, D. Holleville, P. Lemonde, A. Amy-Klein, Ch. Chardonnet, G. Santarelli, "Progress to a full optical link on a telecom network," Proceedings of the the 22nd European Frequency and Time Forum, Toulouse, France, 23-25 April 2008.

22.  D. W. Allan, "Statistics of atomic frequency standards" Proc. IEEE, **54**, 221-230 (1966)

23.  S. T. Dawkins, J. J. McFerran, A. N. Luiten, "Considerations on the measurement of the stability of oscillators with frequency counters", IEEE Trans. Ultrason. Ferroelectr. Freq. Control **54**, 918-925 (2007).

24.  D. W. Allan and J. A. Barnes, "A modified "Allan variance" with increased oscillator characterization ability" in *Proceedings of the Thirty Fifth Annual Frequency Control Symposium* (IEEE, 1981), pp. 470-475.

25.    J. A. Barnes, A. R. Chi, L. S. Cutler, D. J. Healey, D. B. Leeson, T. E. Mcgunigal, J. A. Mullen Jr. , W. L. Smith, R. L. Sydnor , "Characterization of Frequency Stability", IEEE Trans. Instrum. Meas. **20**, 105-120 (1971).

FIGURE CAPTIONS

Fig. 1. Phase noise power spectral density ~~versus Fourier frequency~~ of the stabilized laser (solid line) and of a 1-Hz linewidth white frequency noise laser (dashed line)

Fig. 2. Full scheme of the optical frequency transfer over fiber; AOM: acousto-optic modulator; PD: photodiode; PLL: phase locked loop; PC: polarization controller: OC: optical circulator; RF synth: Radio-frequency synthesizer.

Fig. 3. Phase noise power spectral density ~~versus Fourier frequency~~ of the (a) free running, (b) compensated 86-km link and (c) delayed self-heterodyne beat note. Dashed line represents the phase noise of a 1-Hz linewidth laser.

Fig. 4. Fiber noise rejection for 86-km link, experimental data (blue line) and analytical data with proportional gain equal to 2300 and integrator gain equal to 50 (black dashed line). Red dotted line represents the delay-induced rejection floor.

Fig.5. Fractional frequency instability of the 86-km free running link (blue triangle) and compensated link (black squares), versus averaging time. The measurement frequency bandwidth is 10 Hz.

Fig. 6. Scheme of the 172-km link (EDFA : Erbium Doped Fiber Amplifiers, AOM, PD, see Fig.3.)

Fig. 7. Phase noise power spectral density of the (a) free running, and (b) compensated 172-km link.

Fig. 8. Fractional frequency instability of the 172-km free running link (blue triangle), compensated link with 10-Hz measurement bandwidth (black squares), and with Λ-type counter (green stars). Empty triangles correspond to the experimental set-up floor (see text).

Figure 1

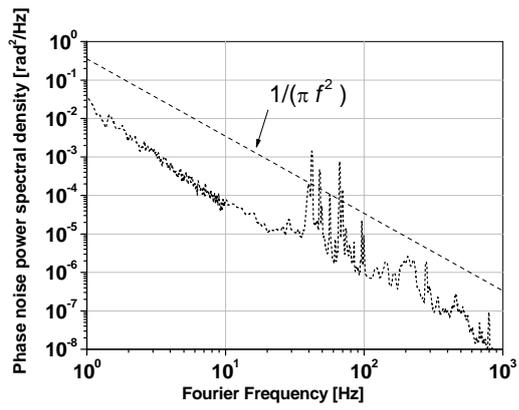

Figure 2

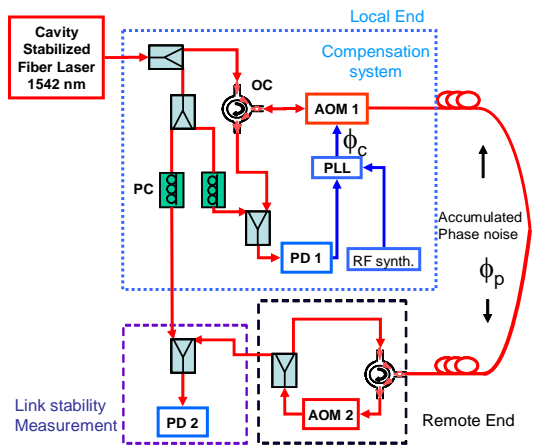

Figure 3

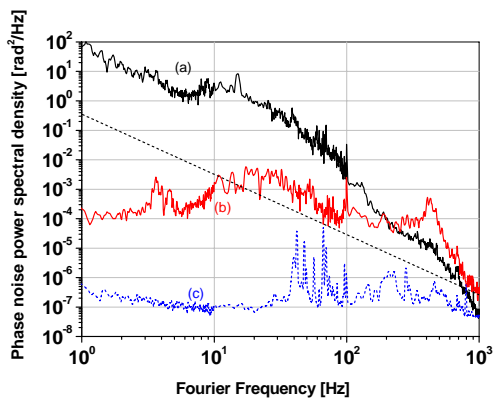

Figure 4

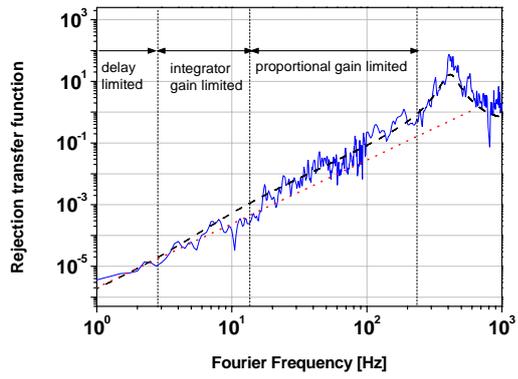

Figure 5

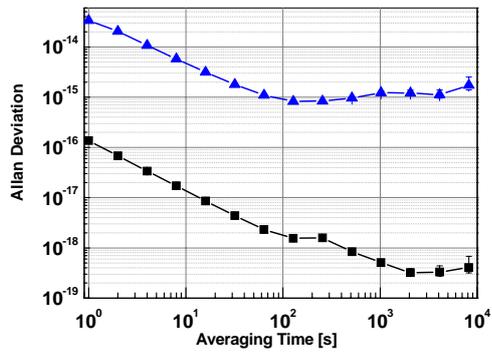

Figure 6

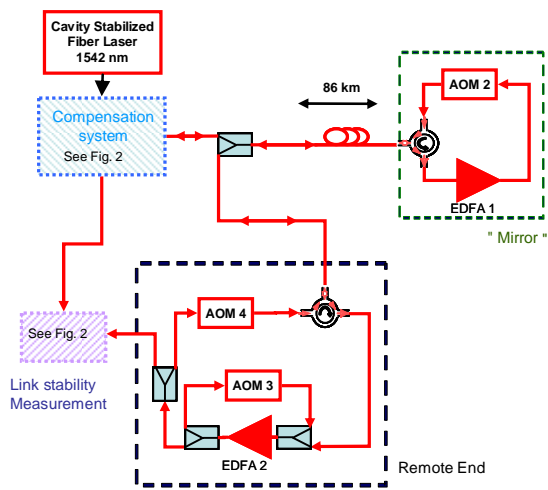

Figure 7

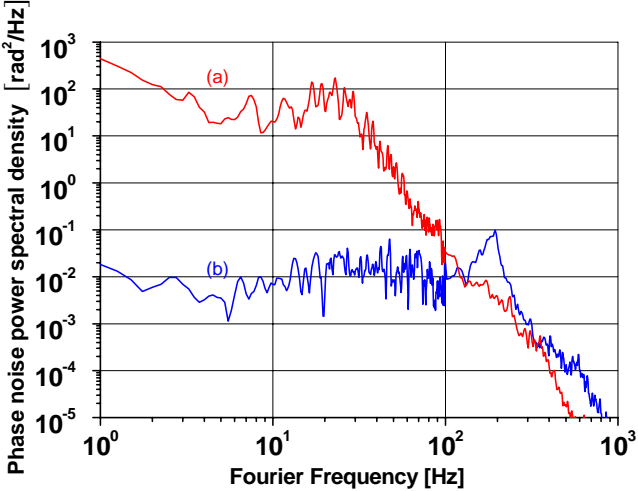

Figure 8

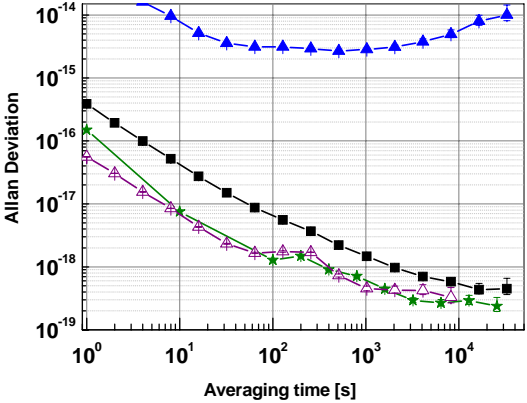